\documentclass[preprint,12pt]{elsarticle}

\usepackage{amssymb}
\usepackage{xcolor}
\usepackage{multirow}%
\usepackage{array}
\usepackage{tabularx}
\usepackage{natbib}
\usepackage{url}
\usepackage{fontawesome}

\journal{IST}

\begin{document}

\begin{frontmatter}



\title{Raising AI Ethics Awareness through an \emph{AI Ethics Quiz} for Software Practitioners}

\author[label 1]{Aastha Pant} 
\author[label 1]{Rashina Hoda}
\author[label 2]{Paul McIntosh}

\affiliation[label 1]{organization={Monash University},
            addressline={Wellington Road}, 
            city={Melbourne},
            postcode={3800}, 
            state={Victoria},
            country={Australia}}
\affiliation[label 2]{organization={RMIT},
            addressline={124 La Trobe St},
            city={Melbourne},
            postcode={3000},
            state={Victoria},
            country={Australia}}

\begin{abstract}

\textbf{Context:} Today, ethical issues surrounding artificial intelligence (AI) systems are increasingly prevalent, highlighting the critical need to integrate AI ethics into system design to prevent societal harm. Raising awareness and fostering a deep understanding of AI ethics among software practitioners is essential for achieving this goal. However, research indicates a significant gap in practitioners' awareness and knowledge of AI ethics and ethical principles. While much effort has been directed toward helping practitioners operationalise AI ethical principles such as \emph{fairness}, \emph{transparency}, \emph{accountability}, and \emph{privacy}, less attention has been paid to raising initial awareness, which should be the foundational step. 

\textbf{Objective:} Addressing this gap, we had developed a software-based tool, the \emph{AI Ethics Quiz}, aimed at raising awareness and enhancing the knowledge of AI ethics among software practitioners. In this study, our objective was to organise interactive workshops, introduce the \emph{AI Ethics Quiz}, and evaluate its effectiveness in enhancing awareness and knowledge of AI ethics and ethical principles among practitioners. 

\textbf{Method:} We conducted two one-hour workshops (one in-person and one online) involving 29 software practitioners. Data was collected through a pre-quiz questionnaire, the \emph{AI Ethics Quiz}, and a post-quiz questionnaire. 

\textbf{Results:} The anonymous responses from the pre-and post-quiz questionnaires revealed that the \emph{AI Ethics Quiz} significantly improved practitioners' awareness and understanding of AI ethics. Additionally, practitioners found the quiz engaging and reported it created a meaningful learning experience regarding AI ethics. In this paper, we share insights gained from conducting these interactive workshops and introducing the \emph{AI Ethics Quiz} to software practitioners. 

\textbf{Conclusion:} We also provide recommendations for software companies and leaders to adopt similar initiatives, which may help them enhance practitioners' awareness and understanding of AI ethics.
\end{abstract}



\begin{keyword}
 AI ethics, AI Ethics Quiz\footnote{https://interactive-ai-ethics-quiz.herokuapp.com/}, awareness, understanding, software practitioners



\end{keyword}

 \end{frontmatter}



\section{Introduction} \label{sec:Introduction}
The importance of ethics in AI-based systems has been highlighted significantly in recent years \cite{hagendorff2020ethics}. Ethics in AI-based systems is more crucial now than ever as these systems are being used across various domains such as health, transport, banking, and recruitment \cite{mehrabi2021survey}. The widespread use of AI-based systems in diverse domains has consequently led to a rise in ethical issues associated with these systems. For instance, people today face various ethical challenges related to AI-based systems, particularly concerning \emph{bias}. Notable examples include Amazon's recruitment tool, which exhibited bias against women by favoring male candidates during the recruitment process \cite{Amazon}, and Apple's credit card, which was biased against women by offering them lower credit limits compared to men \cite{Apple}. Additionally, there have been significant \emph{privacy} concerns, such as Generative AI tools replicating artists' work without consent, prompting artists to use various measures to protect their creations \cite{Art}. Furthermore, AI-based systems have demonstrated issues related to \emph{transparency} and \emph{bias} in image generation \cite{ImageGen}. The growing number of ethical incidents involving AI-based systems underscores the need to raise awareness and enhance the knowledge of software practitioners\footnote{Practitioners who are currently engaged with AI-based systems, had prior experience with such systems, or were planning to work with them in the future.} on AI ethics. Since software practitioners are closely involved with these systems, they are in a pivotal position to make necessary changes and improvements \cite{pant2024ethics}.

\par Research has shown that practitioners often lack awareness of AI ethics and the knowledge of ethical principles related to AI \cite{vakkuri2020just, pant2024ethics}. Various guidelines and frameworks on AI ethics have been introduced in recent years by different tech companies like IBM \cite{IBM}, Microsoft \cite{Microsoft}, and Google \cite{Google}, and by different countries and continents such as Australia \cite{Australisethics} and Europe \cite{EUethics}. These initiatives aim to guide practitioners in developing ethical AI-based systems. However, merely having guidelines has not proven to be effective in enhancing the knowledge of practitioners on software ethics \cite{mcnamara2018does}, and a similar outcome is predicted in the area of AI ethics \cite{vakkuri2020just}. Along with that, over the years, various approaches have been developed and used to aid practitioners in integrating ethics into AI-based system development. Emphasis has been placed on enabling practitioners to operationalise specific ethical principles of AI, such as \emph{fairness} \cite{weerts2023fairlearn, vasudevan2020lift, Playingwith}, \emph{transparency/explainability} \cite{arya2021ai, SHAP, MicrosoftInterpret}, and \emph{accountability/responsibility} \cite{AlgorithmicAccountability, PwC}. However, there has been comparatively limited focus on raising awareness and enhancing knowledge and understanding of AI ethics among practitioners which should be the initial step towards developing ethical AI-based systems. 

\par Therefore, in response to the identified gap in the literature and the significance highlighted we had previously developed an interactive \emph{AI Ethics Quiz} \cite{teo2023would} to raise awareness and improve knowledge of AI ethics among software practitioners. In this study, our primary goal was to introduce the \emph{AI Ethics Quiz} to software practitioners to assess its effectiveness in raising their awareness and enhancing their understanding of AI ethics. Specifically, we aimed to address the following research questions (RQs):

\begin{itemize}
    \item \textbf{Ethical decision-making of practitioners}
    \begin{itemize}
              \item \textbf{RQ1:} To what extent do software practitioners make AI ethical decisions when faced with hypothetical scenarios involving ethical dilemmas?
              \end{itemize}

\item \textbf{Efficacy of the quiz}
              \begin{itemize}
           \item \textbf{RQ2:} How aware were software practitioners of the concept of
AI ethics before taking the \emph{AI Ethics Quiz}? How aware are they after taking the \emph{AI Ethics Quiz}?
    \item \textbf{RQ3:} How good was software practitioners’ knowledge/ under-
standing of the concept of AI ethics before taking the \emph{AI Ethics Quiz}? How good is it after taking the \emph{AI Ethics Quiz}?
    \item \textbf{RQ4:} How confident were software practitioners about approaching ethical scenarios in their practice before taking the \emph{AI Ethics Quiz}? How confident are they after taking the \emph{AI Ethics Quiz}?
    \end{itemize}

\item \textbf{Engagement and usability of the quiz}
\begin{itemize}
    
    \item \textbf{RQ5:} How engaging is the \emph{AI Ethics Quiz} to learn about AI ethics?
    \item \textbf{RQ6:} How difficult is it to find the ideal answers to the questions in the \emph{AI Ethics Quiz}?
    \item \textbf{RQ7:} How well-integrated are the \emph{AI Ethics Quiz} scenarios, questions, and answer options to create a meaningful learning experience about AI ethics?
    \end{itemize}
   \end{itemize}

\par The \emph{AI Ethics Quiz} comprised four hypothetical scenarios, each with three to four questions (13 in total), and had four answer options per question. Detailed information about the design and development of the \emph{AI Ethics Quiz} can be found in the paper by \citet{teo2023would}. 

\par This study aimed to address the research questions by introducing the \emph{AI Ethics Quiz} to 29 software practitioners employed in an IT-based company. The primary objective was to evaluate the quiz's effectiveness in raising awareness and enhancing knowledge of AI ethics and ethical principles among software practitioners.
To achieve this, we organised two one-hour workshops—one in-person and one online with software practitioners. Then, we collected data through a survey to assess the quiz's effectiveness in enhancing their awareness and understanding of AI ethics. The primary contribution of this study is the introduction of the \emph{AI Ethics Quiz} to software practitioners and the evaluation of its effectiveness in raising software practitioners' awareness and understanding of AI ethics and ethical principles. In addition, we share our experiences conducting workshops, engaging with software practitioners on AI ethics, and assessing the effectiveness of the \emph{AI Ethics Quiz}, highlighting lessons learned throughout the process. We also provide recommendations for software companies and leaders to adopt similar initiatives, aiming to enhance practitioners' awareness and understanding of AI ethics.

\section{Background and Related Work} \label{sec:RW}
\par There is no universally accepted definition of the term \emph{AI ethics} \cite{kazim2021high,vakkuri2018key}. In this study, we adopt the definition of \emph{AI ethics} provided by \citet{siau2020artificial}, which states, ``the principles of developing AI to interact with other AIs and humans ethically and function ethically in society." 

\par Today, AI-based systems are used across various domains, including health, transport, banking, and recruitment \cite{mehrabi2021survey}. The proliferation of these systems has also amplified the ethical issues associated with their use. In recent years, various ethical issues have emerged in AI, including biases in AI-driven tools like \emph{Apple's Credit Card}, which discriminated against women by offering them lower credit limits compared to men \cite{Apple}, and a healthcare risk prediction algorithm in the United States, which led to black people being denied essential care because the algorithm equated health needs with financial costs \cite{USAlgo}. Privacy concerns have also arisen with AI applications like GitHub's ML-powered GitHub Copilot, which used copyrighted source code as training data without consent \cite{twitter}, and Slack, which incorporated customer data into its AI models for training purposes \cite{Slack}. Issues related to explainability were also seen in an ML algorithm that was developed to predict the risks of a patient after cardiac surgery \cite{MIT}. All these ethical incidents related to AI-based systems highlight the importance of considering AI ethics during the development of AI-based systems. 

\par Raising awareness and educating people on \emph{AI ethics} has become increasingly critical in today's technology-driven world. For example, there have been efforts to raise awareness and educate students on \emph{AI ethics} who are taking technology courses, and there have been discussions and approaches used recently about incorporating the topic of \emph{AI ethics} into the curriculum \cite{AIclub,dabbagh2024ai}. Moreover, in its recent \emph{Recommendation on the Ethics of AI}, the United Nations Educational, Scientific and Cultural Organization (UNESCO) advocates for comprehensive AI literacy education. UNESCO recommends that Member States educate all stakeholders involved in AI-based systems (users, developers, and others) stating the need to \emph{``provide adequate AI literacy education to the public at all levels in all countries to empower people and mitigate digital divides and access inequalities resulting from the widespread adoption of AI systems"}. Similarly, the Australian Human Rights Commission has also recommended providing comprehensive digital literacy programs and training to not only students but also teachers on the \emph{ethics} of Generative AI and its responsible use \cite{HumanRights}. Given the growing emphasis on the need for widespread awareness and education in \emph{AI ethics}, we are taking a step in that direction.

\par Over the years, various countries, continents, and tech companies have developed ethical guidelines for AI. For instance, Australia has established its own AI ethics guidelines \cite{Australisethics}, Europe has formulated its AI ethics framework \cite{EUethics}, and tech giants like Microsoft \cite{Microsoft}, IBM \cite{IBM}, and Google \cite{Google} have created their own AI ethics principles. These guidelines serve as a guide for practitioners, assisting in the development of ethical AI-based systems. For example, in Australia's AI Ethics Framework \cite{Australisethics}, there is a clear acknowledgment that practitioners should consider questions like \emph{``Will the AI system you are developing or implementing be used to make decisions or in other ways have a significant impact (positive or negative) on people (including marginalised groups), the environment or society?"} and \emph{``Are you unsure about how the AI system may impact your organisation or your customers/ clients?"} if they are developing AI-based systems. This underscores the critical need for practitioners to be well-versed in AI ethics and ethical principles when developing AI-based systems.

\par Empirical research has shown that practitioners often lack both awareness and a deep understanding of AI ethics and ethical principles \cite{vakkuri2020just,pant2024ethics}. For instance, our previous empirical study showed that only 13\% of AI practitioners were extremely aware of the concept of AI ethics. Likewise, an empirical study by \citet{mcnamara2018does} found that the ethical guidelines issued by the Association for Computing Machinery (ACM) had minimal impact on software developers. Despite the guidelines, developers continued their previous practices, suggesting a significant gap in ethical knowledge among them. As a result, the Ethically Aligned Design (EAD) guidelines acknowledge similar ethical issues may arise when it comes to AI ethics \cite{vakkuri2020just}. On the other hand, empirical studies have also shown that practitioners are aware of some specific ethical principles of AI such as \emph{transparency} \cite{vakkuri2019ethically, mark2019ethics}, \emph{fairness} \cite{holstein2019improving}, \emph{accountability/ responsibility} \cite{veale2018fairness, rothenberger2019relevance}, and \emph{privacy} \cite{ibanez2022operationalising, ryan2021research, pant2024ethics}. This underscores the critical need and importance of enhancing practitioners’ awareness and understanding of AI ethics and ethical principles. 

\par Education can enhance awareness and identify potentially morally critical situations, and it is recommended that software practitioners receive this training as part of their formal education, integrated into university curricula \cite{leonelli2016locating, garzcarek2019approaching, burton2017ethical, martin2020ethical}. However, as the topic of \emph{AI ethics} is new, many practitioners may not have received this education during their formal training. Providing ethical training to software practitioners and conducting regular group evaluations of methods, strategies, and tools are crucial for fostering ongoing ethical awareness and collaborative learning \cite{leonelli2016locating}. 

\par Significant focus has been placed on assisting practitioners in operationalising specific ethical principles of AI such as \emph{fairness}, \emph{transparency}, \emph{accountability}, and \emph{privacy}. Various tools have been developed to aid practitioners in these efforts. For example, tools like \emph{Fairlearn} \cite{weerts2023fairlearn}, \emph{Playing with AI Fairness} \cite{Playingwith}, \emph{LinkedIn Fairness Toolkit (LiFT)} \cite{vasudevan2020lift}, and \emph{Audit AI} \cite{AuditAI} have been created to help develop \emph{fair} AI-based systems. Similarly, tools like \emph{AI Explainability 360 Toolkit} \cite{arya2021ai}, \emph{SHAP} \cite{SHAP}, and \emph{Microsoft InterPret} \cite{MicrosoftInterpret} assist in operationalising the \emph{explainability} principle of AI. Additionally, tools such as the \emph{Algorithmic Accountability Policy Toolkit} \cite{AlgorithmicAccountability} and \emph{PwC’s Responsible AI} \cite{PwC} have been developed to help practitioners implement the \emph{accountability} principle of AI. However, limited tools and methods have been developed to raise awareness of AI ethics among practitioners. For example, a software-based tool namely \emph{Moral Machine} was developed with the aim of raising awareness of AI ethics among software practitioners \cite{awad2018moral}. Nonetheless, the effectiveness and external validity of these tools and methods remain uncertain \cite{morley2021ethics}. Therefore, we developed an \emph{AI Ethics Quiz} \cite{teo2023would} and introduced it to software practitioners to assess if it enhanced their awareness and understanding of AI ethics. We conducted both in-person and online workshops, which included a talk on AI ethics, followed by administering the \emph{AI Ethics Quiz}. This combination of a talk and an interactive quiz in a workshop was intentionally designed to create an engaging and interactive learning experience for practitioners, as didactic approaches alone often lack engagement and can be ineffective \cite{lane2015new}. The details about the methodology used in this study are presented next.

\section{Methodology} \label{sec:Methodology}
\begin{figure}[h!]
    \centering
    \includegraphics[scale=0.40]{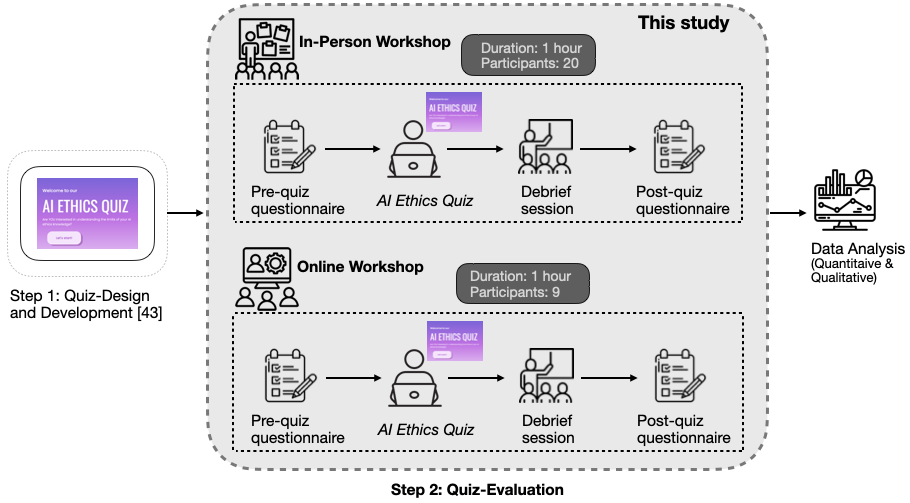}
    \caption{Research methodology of our study (Step 1 was conducted in a preceding study \cite{teo2023would} while Step 2 is presented in this paper)}
    \label{fig:Methodology}
\end{figure}


\subsection{AI Ethics Quiz- Design and Development} \label{sec:Quiz_design}
We had developed an interactive \emph{AI Ethics Quiz} as a web application to help AI practitioners self-assess their knowledge of AI ethics and ethical principles \cite{teo2023would}. The quiz includes hypothetical scenarios with accompanying questions, providing four answer options for each question. There are a total of four scenarios, each containing three to four questions (13 questions in total). These questions present users with varying degrees of ethical dilemmas. Users can choose an answer from four options, each labeled as `least desirable', `bearable', `less than ideal', and `ideal', corresponding to scores of 1, 2, 3, and 4, respectively. The final score is calculated by summing the scores of all their answers. Although the final score is not directly visible to users, the quiz contains a \emph{Result Summary} page at the end that shows the feedback to users on their overall performance. The feedback is categorised into four levels based on the final score. A score of 76-100\% is labeled as an ``Excellent attempt", with participants recognised as \emph{AI ethics experts}. Those who score 51-75\% receive a ``Very good attempt" rating, indicating they are on track to becoming AI ethics experts. Participants with a score of 26-50\% are given a ``Satisfactory attempt" rating, suggesting that with a bit more knowledge, they can achieve expertise in AI ethics. Finally, a score of 0-25\% is marked as an ``Unsatisfactory attempt", highlighting significant room for improvement \cite{teo2023would}. Additionally, users can review a summary of their results for each question and receive feedback on their responses. 

\par After designing the quiz, it was initially refined through a preliminary study involving 50 students enrolled in the Bachelor of Software Engineering course at Monash University, undertaking the Software Engineering Research Project unit. This initial trial and refinement were focused solely on assessing the technical functionality of the quiz and identifying areas for improvement, whereas, the objective of this study was to introduce the quiz to real-world software practitioners to determine its effectiveness in enhancing their awareness and understanding of AI ethics. A detailed description of the design and development of the \emph{AI Ethics Quiz} and its preliminary trial and refinement is available at \citet{teo2023would}. Figure \ref{fig:Methodology} shows the methodology used in our study.

\subsection{AI Ethics Quiz- Use and Evaluation}
We introduced our \emph{AI Ethics Quiz} to 29 software practitioners of an Australian IT company who were either currently working with AI-based systems, had experience with such systems, or were planning to work with them in the future.

\subsubsection{Ethical research considerations}
Given that one of the authors was both a team leader at the company and part of the workshop's organising team, it was important to address potential conflicts of interest. Therefore, ethics approval was obtained from the Human Research Ethics Committee (HREC) at Monash University (approval number: 35521) as the first step. Following this, the author advertised this research by circulating a flyer among their team, associates, and colleagues within the company using an online portfolio booking system. The explanatory statement of the research was also shared which provided details about the workshop, such as the date, time, and venue, and included a QR code for registration. The flyer also noted that attendees should bring their laptops. Additionally, it clarified that attendees could participate in the workshop regardless of whether they chose to participate in the anonymous research study. The purpose of doing this was to attract the maximum number of participants for the workshop. Along with that, the explanatory statement also highlighted the anonymous nature of the data collection process (through the quiz and the pre/post quiz questionnaires), participants' discretion in having their anonymous data included in the research, and their voluntary participation. These standard measures of ethical research were taken to ensure participants' privacy and their learning experience through the workshop. For instance, we repeatedly informed participants that they could attend the workshop regardless of their decision to join the anonymous research study. This clarification was intended to encourage equal participation in the learning opportunity provided by the workshop while only interested candidates participated in the anonymous research study. All data collected from the quiz and pre/post-quiz questionnaires were kept anonymous. Additionally, participants were informed that one of the authors, who served as one of the team leaders at the company, did not have access to the raw data to avoid any conflict of interest. The workshop participants primarily consisted of a small group from the author's (one of the team leaders) team, with the majority being their colleagues and associates within the company.

\subsubsection{Workshop context}
After obtaining ethics approval from the HREC, we aimed to use our \emph{AI Ethics Quiz} to raise awareness and understanding about AI ethics among software practitioners. As previously mentioned, the workshop participants were software practitioners working in an IT company based in Australia. Participants in the workshop either worked with AI-based systems, had experience with AI-based systems development, or were willing to work with AI-based systems in the future. We did not collect demographic information from workshop attendees but only from those who participated in the anonymous research study.

\subsubsection{AI ethics workshop} \label{Sec:AIethics_workshop}
A one-hour workshop was conducted twice- one in-person and one online. 

\textbf{In-person workshop:} 30 employees registered for the in-person workshop. Out of 30, 23 participants attended the workshop and took the \emph{AI Ethics Quiz}, and out of those 23 attendees, 20 participated in the anonymous research study (part of the in-person workshop). 

 \textbf{Online workshop:} Based on the positive feedback from the in-person workshop, we decided to conduct a second workshop online after one month. 18 employees registered for the online workshop, of which 11 participants attended and took the \emph{AI Ethics Quiz}. Out of 11 participants in the online workshop, 9 participated in the anonymous research study (part of the online workshop) making the total number of participants 29. The details of the anonymous research study are described in the next section.

 Both the in-person and online workshops had the same format. The first 20 minutes in both workshops were dedicated to delivering a talk on \emph{AI ethics}. After that, we discussed the workshop agenda and briefed on the data collection process for the anonymous research study. The data collection was conducted in three steps: (i) completing the pre-quiz questionnaire, (ii) taking the \emph{AI Ethics Quiz}, and (iii) completing the post-quiz questionnaire. Those 29 participants completed the pre-quiz questionnaire, took the \emph{AI Ethics Quiz}, and completed the post-quiz questionnaire.

\subsubsection*{ \textbf{Pre-quiz questionnaire:}}
In the in-person workshop, the talk on \emph{AI ethics} was delivered by two of the authors. The talk was designed to familiarise participants with the concept of AI ethics, highlight its significance in today's time, and provide a brief overview of Australia's AI Ethics Principles \cite{Australisethics}. After the talk, the data collection process for the anonymous research study began. The first step for the participants was to complete the pre-quiz questionnaire which was created using \emph{Qualtrics}. However, before that, to make the participation anonymous and track the responses of the participants, participants were required to pick a random number from a bowl that served as their ID. 
Only those interested in participating in the research study, i.e., 20 out of 23, completed the pre-quiz questionnaire in the in-person workshop.

\par In the online workshop, the talk on \emph{AI ethics} was delivered by one of the authors, followed by the data collection process for the anonymous research study. The workshop was conducted on \emph{Zoom}. As in the in-person session, the first step of the data collection process was taking the pre-quiz questionnaire. However, to ensure anonymous participation and track the responses, participants were asked to think of a number between 300 and 700 and use it as their ID. 
Only those interested in participating in the research study, i.e., 9 out of 11, completed the pre-quiz questionnaire in the online workshop.

\par The pre-quiz questionnaire contained two types of questions. The first type consisted of demographic questions such as age, gender, education, job title, AI/ML activities they were involved in, and years of experience. The purpose of these demographic questions was to gather background information about the participants. To ensure anonymity, no personally identifiable information, such as names or email addresses, was collected. Table \ref{Tab:Demo} shows the demographics of the participants. Among the 29 participants, 22 were men, 6 were women, and 1 preferred not to disclose their gender. The largest age groups were 30 to 39 years and 50+ years, each representing 10 participants, while 4 participants fell within the 20 to 29-year age group. Additionally, the majority (18 out of 29) had no experience in AI/ML development. Of the remaining participants, 9 had up to 5 years of experience, and 2 had between 5 to 10 years of experience in developing AI-based systems. The demographic questions are provided in \ref{appendix1}.

\par The second type of questions gathered information about the participants' awareness and understanding of AI ethics. For example, participants were asked to rate their awareness and understanding of AI ethics on a scale of 1 to 5, describe their preferred ways of learning about AI ethics, indicate their awareness of AI ethics principles, and express their confidence in making ethical decisions regarding AI technologies. The purpose of these questions was to gain an overall view of the participant's knowledge and understanding of AI ethics before introducing our \emph{AI Ethics Quiz} to them.

\begin{table}[ht]
\centering
\caption{Demographics of the Participants} \label{Tab:Demo}     
\scriptsize
\begin{tabular} {>{\raggedright\arraybackslash}p{0.7cm}>{\raggedright\arraybackslash}p{1.2cm}>{\raggedright\arraybackslash}p{2.2cm}>{\raggedright\arraybackslash}p{2.5cm}>{\raggedright\arraybackslash}p{4.8cm}}

 \hline\noalign{\smallskip}
P\_Id & Age Range (years) & Gender & Education & Job Title  
\\

\hline\noalign{\smallskip}

P1 & 30-39 & Man & Bachelor &  Software developer 
\\

P2 & 30-39 & Man & Master &  Software engineer 
\\

P3 & 20-29 & Man & Bachelor &  Business analyst 
\\

P4 & 30-39 & Man  & Master &  Business analyst 
\\

P5 & 20-29 & Man  & High school &  Software developer 
\\

P6 & 20-29 & Man  & High school & Software developer 
\\
P7 & 50+ & Man  & Bachelor & Manager 
\\

P8 & 30-39 & Man  & Master & Software development manager 
\\

P9 & 50+ & Prefer not to say  & Ph.D. or higher & Data scientist 
\\

P10 & 50+ & Man  & Master & Software engineer 
\\

P11 & 30-39 & Man  & Bachelor & Software engineer 
\\

P12 & 30-39 & Man  & Master & Analyst 
\\

P13 & 50+ & Man  & Ph.D. or higher & Prefer not to say 
\\

P14 & 40-49 & Man  & Master & Lead developer 
\\

P15 & 50+ & Man  & Bachelor & Enterprise architect 
\\

P16 & 40-49 & Woman  & Bachelor & Analyst 
\\

P17 & 40-49 & Man  & Bachelor & Software developer 
\\

P18 & 20-29 & Man  & Bachelor & Senior full stack web developer 
\\

P19 & 30-39 & Woman  & Bachelor & Business analyst 
\\

P20 & 30-39 & Man  & Master & Software engineer 
\\

P21 & 30-39 & Man  & Master & Business Analyst 
\\

P22 & 50+ & Woman  & Master & Analyst 
\\

P23 & 40-49 & Man & Bachelor & Content writer for the web \\

P24 & 40-49 & Woman & Bachelor & Capability Lead \\

P25 & 50+ & Man & Bachelor & Lead developer \\

P26 & 50+ & Woman & Bachelor & Analyst  \\

P27 & 30-39 & Man & Bachelor & Senior engineer \\

P28 & 50+ & Woman & High school & Software developer \\

P29 & 50+ & Man & Master & Cyber Security Professional \\
\noalign{\smallskip}\hline
\end{tabular}
\end{table}

\subsubsection*{\textbf{Individual AI Ethics Quiz}}
The second step of the data collection process was completing the \emph{AI Ethics Quiz}. In the in-person workshop, the link to the quiz was shared on the screen, while in the online workshop, it was shared via chat. In both workshops, participants were reminded to enter their ID before taking the \emph{AI Ethics Quiz}. Then, they were instructed to complete the quiz independently, without discussing it with fellow participants during the workshop. This was done to ensure that participant's responses to the quiz were not influenced by others. 
They were reminded that a collaborative debrief session would follow once everyone completed the quiz. Participants were also reminded of the anonymous nature of the quiz and that participation was voluntary. 15 minutes were allotted to complete the quiz in both workshops. Participants could view their overall performance feedback after completing the quiz. They were asked not to see the feedback for each question. This was intentionally done to encourage discussion about the ideal answers and their reasoning during the debriefing session with all participants.

\subsubsection*{\textbf{Collaborative debrief}}
\begin{figure}[h!]
    \centering
    \includegraphics[scale=0.25]{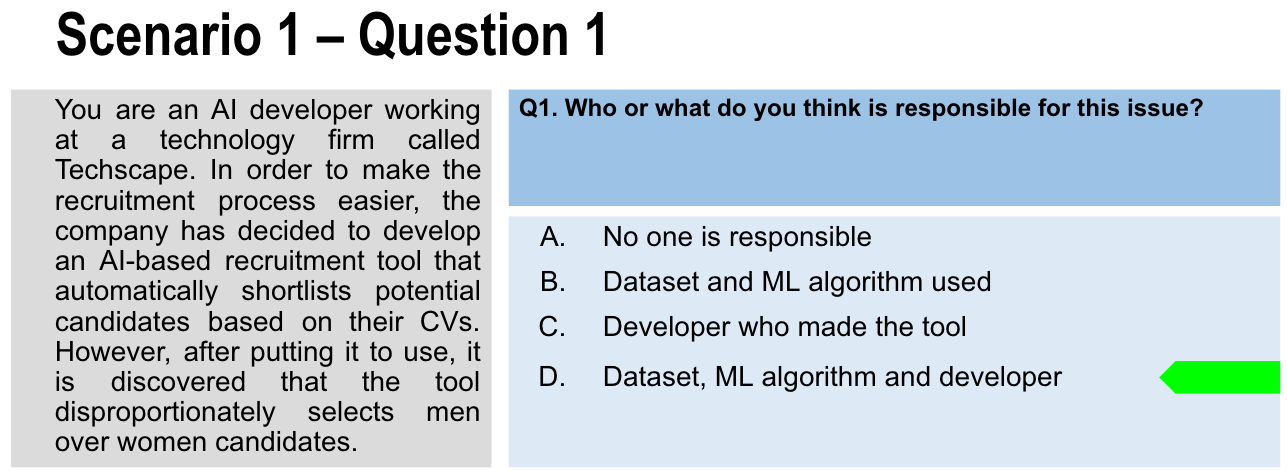}
    \caption{Scenario 1-Q1, along with its answer options and an animated green tag indicating the ideal answer}
    \label{fig:Scenario}
\end{figure}
Once all participants completed the quiz, the next set of slides was presented in both in-person and online workshops. Each slide displayed a scenario and its corresponding question with four answer options, illustrated in Figure \ref{fig:Scenario} for \emph{Scenario 1- Question 1} of the quiz. Participants were then prompted to share their perspectives on the ideal answer and provide their justifications. 
Differences in opinions were openly discussed, with a reminder of psychological safety and non-judgmental attitudes toward answers. After participants shared and sometimes debated their viewpoints, the instructor revealed the ideal response (animated green tag) and explained why it was ideal given the scenario, referencing relevant codes of ethics. Each of the other answer options was then discussed in turn, highlighting why they were considered, `less than ideal’ `bearable’, or `least desirable’. This process sparked additional questions from participants and fostered an engaging discussion. 

\subsubsection*{\textbf{Post-quiz questionnaire}}
The final step for participants was to complete a post-quiz questionnaire created using \emph{Qualtrics}. This questionnaire aimed to gather anonymous feedback on the effectiveness of the \emph{AI Ethics Quiz} in enhancing their awareness and understanding of AI ethics. A QR code to the post-quiz questionnaire was shared on the screen in both workshops, and participants were reminded to use their ID at the beginning of the questionnaire. All 29 participants completed it within an allocated five-minute time frame.

\par The post-quiz questionnaire investigated whether the quiz helped practitioners enhance their awareness and understanding of AI ethics. Participants rated the quiz's efficacy, engagement level, difficulty in identifying ideal answers, integration of scenarios, questions, and answer options, and their confidence in approaching ethical scenarios after taking the quiz. An open-ended question was included for additional feedback or suggestions to improve the \emph{AI Ethics Quiz}. Table \ref{Tab:Pre/post-quiz} provides a detailed list of all pre and post-quiz questionnaires and answer options.

\begin{table}[h!]
\centering
\caption{Pre/Post-Quiz Questionnaires} \label{Tab:Pre/post-quiz}     
\scriptsize

\begin{tabular} {>{\raggedright\arraybackslash}p{0.3cm}>{\raggedright\arraybackslash}p{6.1cm}>{\raggedright\arraybackslash}p{6.1cm}}

 \hline\noalign{\smallskip}
S.N. & Pre-Quiz Questionnaire & Answer Options  
\\

\hline\noalign{\smallskip}

1 & I am aware of the concept of AI ethics. & 1- Strongly Disagree, 2- Somewhat Disagree, 3- Neither Agree nor Disagree, 4- Somewhat Agree, 5- Strongly Agree
\\

2 & I have a good knowledge/ understanding of the concept of AI Ethics. &  1- Strongly Disagree, 2- Somewhat Disagree, 3- Neither Agree nor Disagree, 4- Somewhat Agree, 5-Strongly Agree
\\

3 & Of all the following ways of learning about AI ethics, please select all the ones that you find engaging. & Textbooks, Quiz, Research articles, Podcasts and videos, Others
\\

4 & Which of the following AI ethical principles are you aware of? Select all that apply. These are a selected list of the majority of the principles considered around the world. (Australia’s AI Ethics Principles) \cite{Australisethics} & Accountability, Contestability, Fairness, Human-centered values, Human, societal, and environmental well-being, Privacy protection and security, Reliability and safety, Transparency and explainability, All, None, Others 
\\

5 & I am confident in making ethical decisions related to AI technologies. & 1- Strongly Disagree, 2- Somewhat Disagree, 3- Neither Agree nor Disagree, 4- Somewhat Agree, 5-Strongly Agree
\\
\hline\noalign{\smallskip}
S.N. & Post-Quiz Questionnaire & Answer Options  
\\

\hline\noalign{\smallskip}

1 & Taking the \emph{AI Ethics Quiz} helped improve my awareness of the concept of AI ethics. & 1- Strongly Disagree, 2- Somewhat Disagree, 3- Neither Agree nor Disagree, 4- Somewhat Agree, 5- Strongly Agree
\\

2 & Taking the \emph{AI Ethics Quiz} helped improve my understanding of the concept of AI ethics. &  1- Strongly Disagree, 2- Somewhat Disagree, 3- Neither Agree nor Disagree, 4- Somewhat Agree, 5- Strongly Agree
\\

3 & The \emph{AI Ethics Quiz} was an engaging way to learn about AI ethics. & 1- Strongly Disagree, 2- Somewhat Disagree, 3- Neither Agree nor Disagree, 4- Somewhat Agree, 5-Strongly Agree
\\

4 & I found it hard to identify the most ideal answers to the questions in the \emph{AI Ethics Quiz}. & 1- Strongly Disagree, 2- Somewhat Disagree, 3- Neither Agree nor Disagree, 4- Somewhat Agree, 5- Strongly Agree
\\

5 & The quiz scenarios, questions, and answer options were well integrated to create a
meaningful learning experience about AI ethics. & 1- Strongly Disagree, 2- Somewhat Disagree, 3- Neither Agree nor Disagree, 4- Somewhat Agree, 5- Strongly Agree
\\

6 & I feel more confident about approaching ethical scenarios in my practice after taking this Quiz. & 1- Strongly Disagree, 2- Somewhat Disagree, 3- Neither Agree nor Disagree, 4-Somewhat Agree, 5- Strongly Agree
\\

7 & Please share any other feedback or suggestions you have about the \emph{AI Ethics Quiz}. & Open-text answer\\

\noalign{\smallskip}\hline

\end{tabular}
\end{table}

\section{Findings} \label{sec:Findings}
\subsection{AI Ethics Quiz Responses}
\subsubsection{Ethical decision-making of practitioners}
\textbf{RQ1:} To what extent do software practitioners make AI ethical decisions when faced with hypothetical scenarios involving ethical dilemmas?
\begin{figure}[h!]
    \centering
    \includegraphics[scale=0.40]{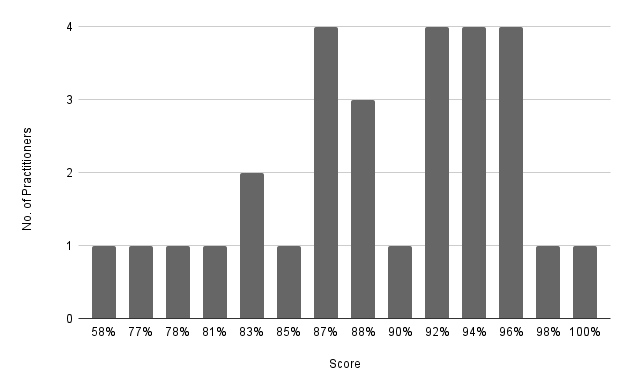}
    \caption{Distribution of quiz scores among 29 practitioners}
    \label{fig:Quiz_Score}
\end{figure}

\par The mean quiz score for answering all 13 questions was 46.17 out of 52 (88.78\%). The highest score achieved by any practitioner was 52 out of 52 (100\%), attained by one participant, while the lowest score was 30 out of 52 (58\%), also attained by one participant. This suggests that most practitioners who took the quiz had a good understanding of AI ethics, as the majority scored above 75\%. In our quiz, a score of 76-100\% is considered an excellent attempt, identifying participants as \emph{AI ethics experts}, as detailed in Section \ref{sec:Quiz_design}. The distribution of scores, ranging from 58\% to 100\%, is shown in Figure \ref{fig:Quiz_Score}.

\subsection{Pre/Post-Quiz Questionnaire Responses}
\par The objective of the pre/post-quiz questionnaire was to get participants' responses regarding the efficacy, engagement, and usability of \emph{AI Ethics Quiz} in enhancing their awareness, and understanding of AI ethics. As mentioned in  Section \ref{Sec:AIethics_workshop}, we obtained complete responses for the pre and post-quiz questionnaires from a total of 29 participants. 

\par
\begin{figure}[h!]
    \centering
    \includegraphics[scale=0.40]{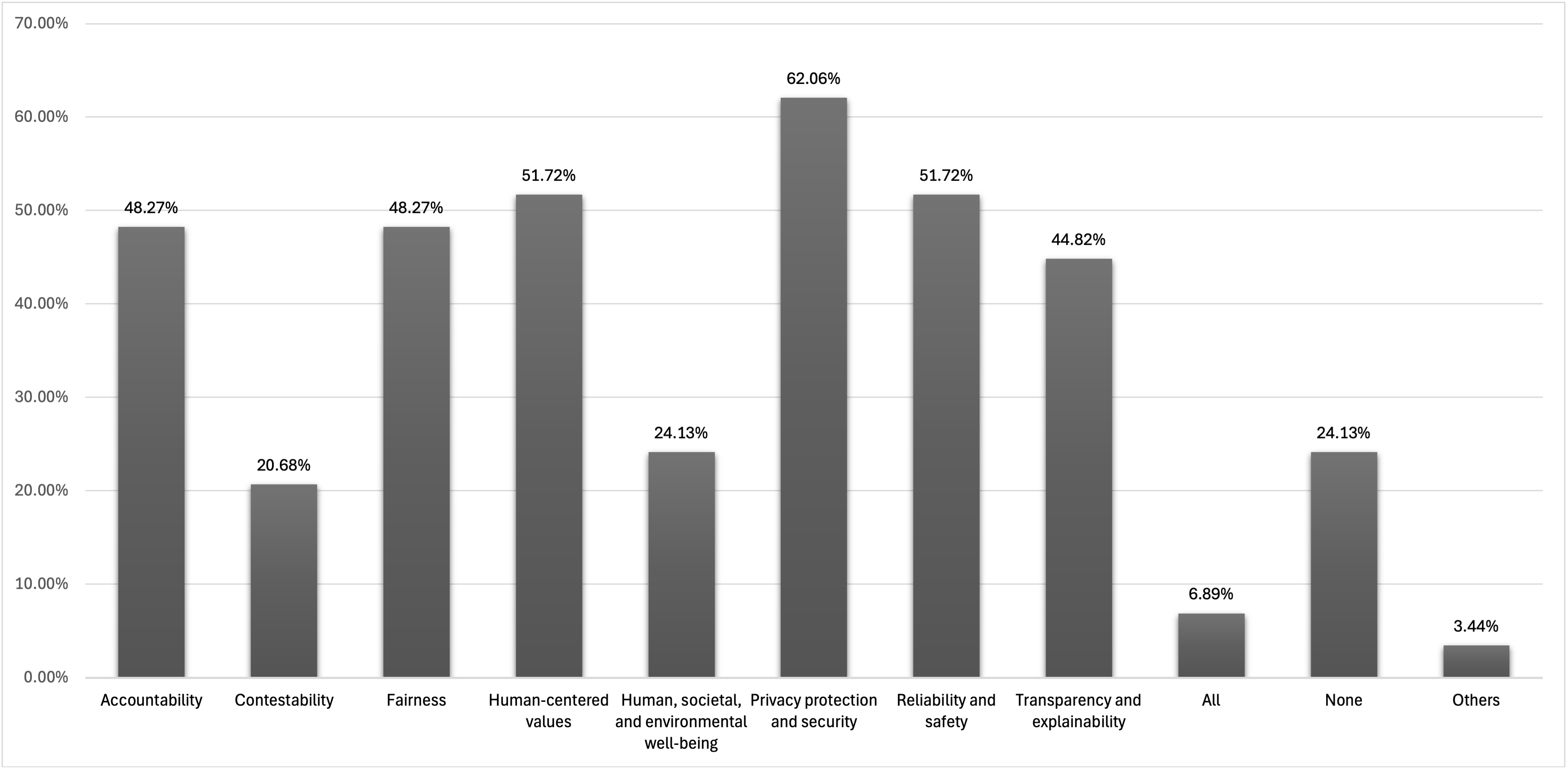}
    \caption{Practitioners' awareness of Australia's AI Ethics Principles \cite{Australisethics} (pre-quiz)}
    \label{fig:Eth_Principle}
\end{figure}

 There were five questions in the pre-quiz questionnaire form out of which three questions provided a Likert-Scale (1= Strongly disagree, 5= Strongly agree), and the remaining two allowed participants to select multiple answers. Before asking participants about their experience with the quiz, we gathered general information about their awareness of AI ethical principles using Australia's AI Ethics Principles List \cite{Australisethics} and the resources they use to learn about AI ethics. Our findings indicate that the majority of participants (62.06\%) were aware of the \emph{`Privacy protection and security'} principle, followed by principles like \emph{`Reliability and safety'} and \emph{`Human-centered values'}, each recognised by 51.72\% of participants. Only 6.89\% of participants were aware of all ethical principles, while 24.13\% was not aware of any AI ethical principles, as shown in Figure \ref{fig:Eth_Principle}.

\begin{figure}[h!]
    \centering
    \includegraphics[scale=0.40]{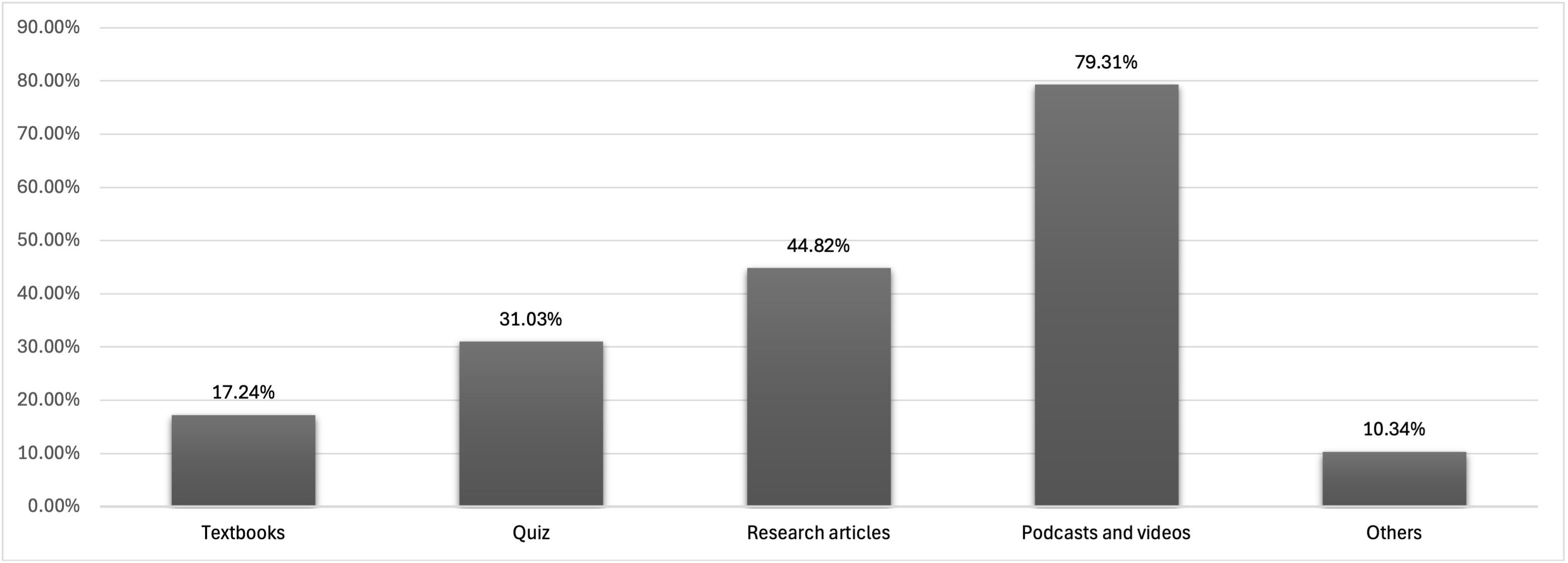}
    \caption{Practitioners' ways of learning about AI ethics (pre-quiz)}
    \label{fig:Sources}
\end{figure}

Similarly, our findings show that the majority of participants (79.31\%) indicated \emph{`podcasts and videos'} as their primary method of learning about AI ethics, followed by \emph{`research articles'} (44.82\%), as shown in Figure \ref{fig:Sources}. Only 17.24\% of participants mentioned \emph{`textbooks'} as their preferred source for learning about AI ethics, while 10.34\% chose the \emph{`others'} option. Participants who selected \emph{`others'} mentioned sources such as \emph{`presentations'} and \emph{`lectures'} at universities to learn about AI ethics. The rest of the findings report on all the remaining three questions of the pre-quiz questionnaire containing predefined answers analysed using descriptive statistics. 

Similarly, the post-quiz questionnaire consisted of seven questions out of which six provided a Likert-Scale (1=Strongly disagree to 5=Strongly agree), while the final question was open-ended. The quantitative data was analysed using descriptive statistics. The findings report on all six questions with predefined answers, whereas the responses to the open-ended question are discussed in the `Discussion' section (Section \ref{sec:Discussion}).

\subsubsection{Efficacy of the quiz:}
\begin{figure}[htpb]
    \centering
    \includegraphics[scale=0.45]{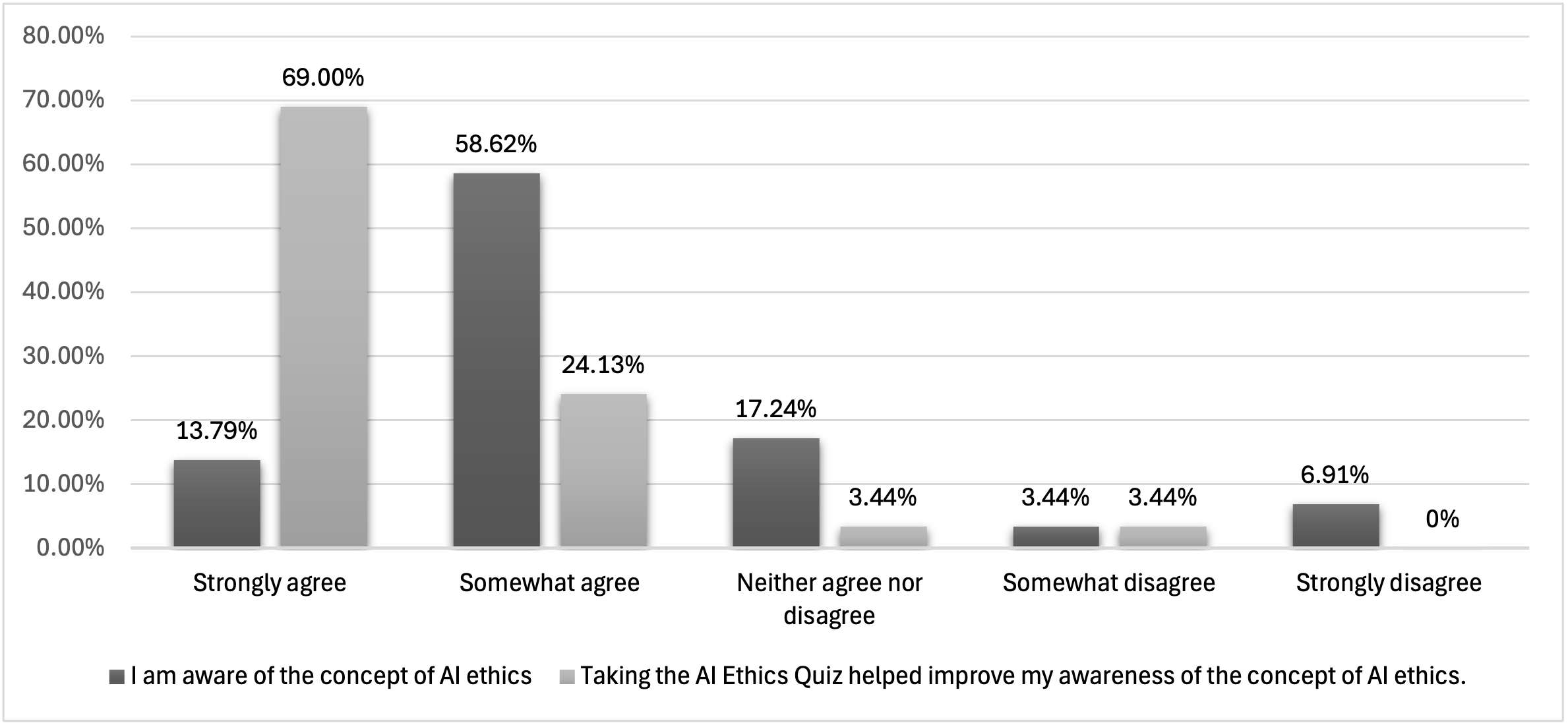}
    \caption{Practitioner responses about the efficacy of the quiz in raising AI ethics awareness (pre vs post-quiz)}
    \label{fig:Awareness}
\end{figure}
\textbf{RQ2:} \textbf{How aware were software practitioners of the concept of AI ethics before taking the \emph{AI Ethics Quiz}? How aware are they after taking the \emph{AI Ethics Quiz}?}

\par A majority (72.41\%) of the practitioners agreed or strongly agreed that they were aware of the concept of AI ethics before taking the \emph{AI Ethics Quiz} as shown in Figure \ref{fig:Awareness}. While 17.24\% of the practitioners neither agreed nor disagreed, 10.35\% disagreed or strongly disagreed with the statement that they were aware of the concept of AI ethics before taking the quiz. Likewise, most (93.13\%) practitioners agreed or strongly agreed that taking the \emph{AI Ethics Quiz} helped them improve their awareness of AI ethics. Only 3.44\% of the practitioners neither agreed nor disagreed, and another 3.44\% disagreed that taking the quiz helped improve their awareness of AI ethics.
Although we did not ask the practitioners to explicitly compare the two, the perceived improvement in awareness of AI ethics was quite higher (by 20.72 percentage points) after taking the quiz compared to before.

\textbf{RQ3:} \textbf{How good was software practitioners' knowledge/ understanding of the concept of AI ethics before taking the \emph{AI Ethics Quiz}? How good is it after taking the \emph{AI Ethics Quiz}?} \label{sec:RQ3}
\begin{figure}[htpb]
    \centering
    \includegraphics[scale=0.45]{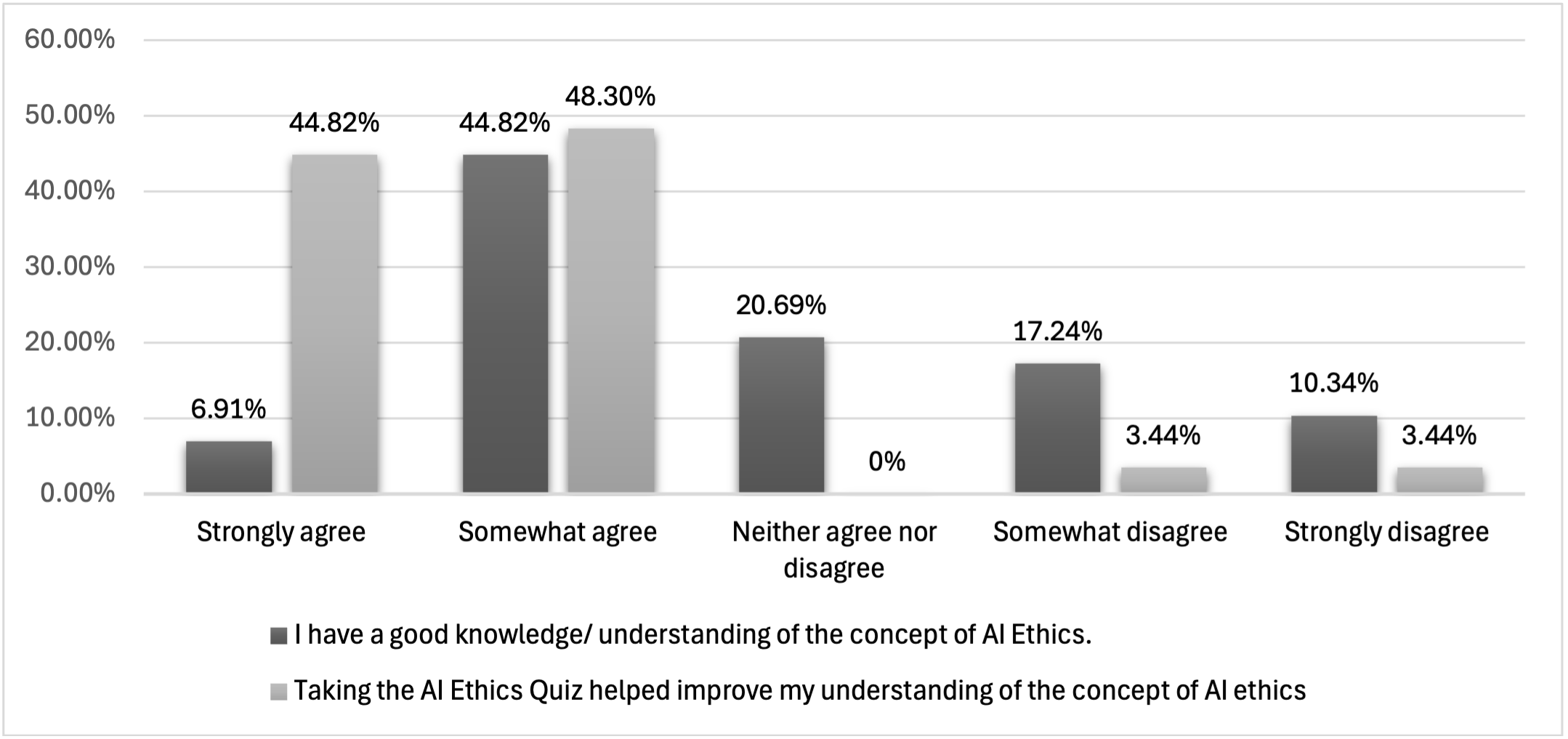}
    \caption{Practitioner responses about the efficacy of the quiz in improving AI ethics knowledge/understanding (pre vs post-quiz)}
    \label{fig:Knowledge}
\end{figure}

\par A majority (51.73\%) of the practitioners agreed or strongly agreed that
they had a good knowledge/understanding of the concept of AI ethics before taking the \emph{AI Ethics Quiz} as shown in Figure \ref{fig:Knowledge}. 20.69\% of the practitioners neither agreed nor disagreed with the statement, whereas, 27.58\% disagreed or strongly disagreed that they had a good knowledge/ understanding of the concept of AI ethics before taking the quiz. Similarly, most (93.12\%) practitioners agreed or strongly agreed that taking the \emph{AI Ethics Quiz} helped them improve their knowledge/ understanding of AI ethics and only 6.88\% disagreed or strongly disagreed that taking the quiz helped improve their knowledge/ understanding of AI ethics. Practitioners reported a significantly greater improvement in their knowledge/ understanding of AI ethics after taking the quiz, with an increase of 41.39 percentage points compared to before.

\textbf{RQ4:} \textbf{How confident were software practitioners about approaching ethical scenarios in their practice before taking the \emph{AI Ethics Quiz}? How confident are they after taking the \emph{AI Ethics Quiz}?}

\begin{figure}[htpb]
    \centering
    \includegraphics[scale=0.45]{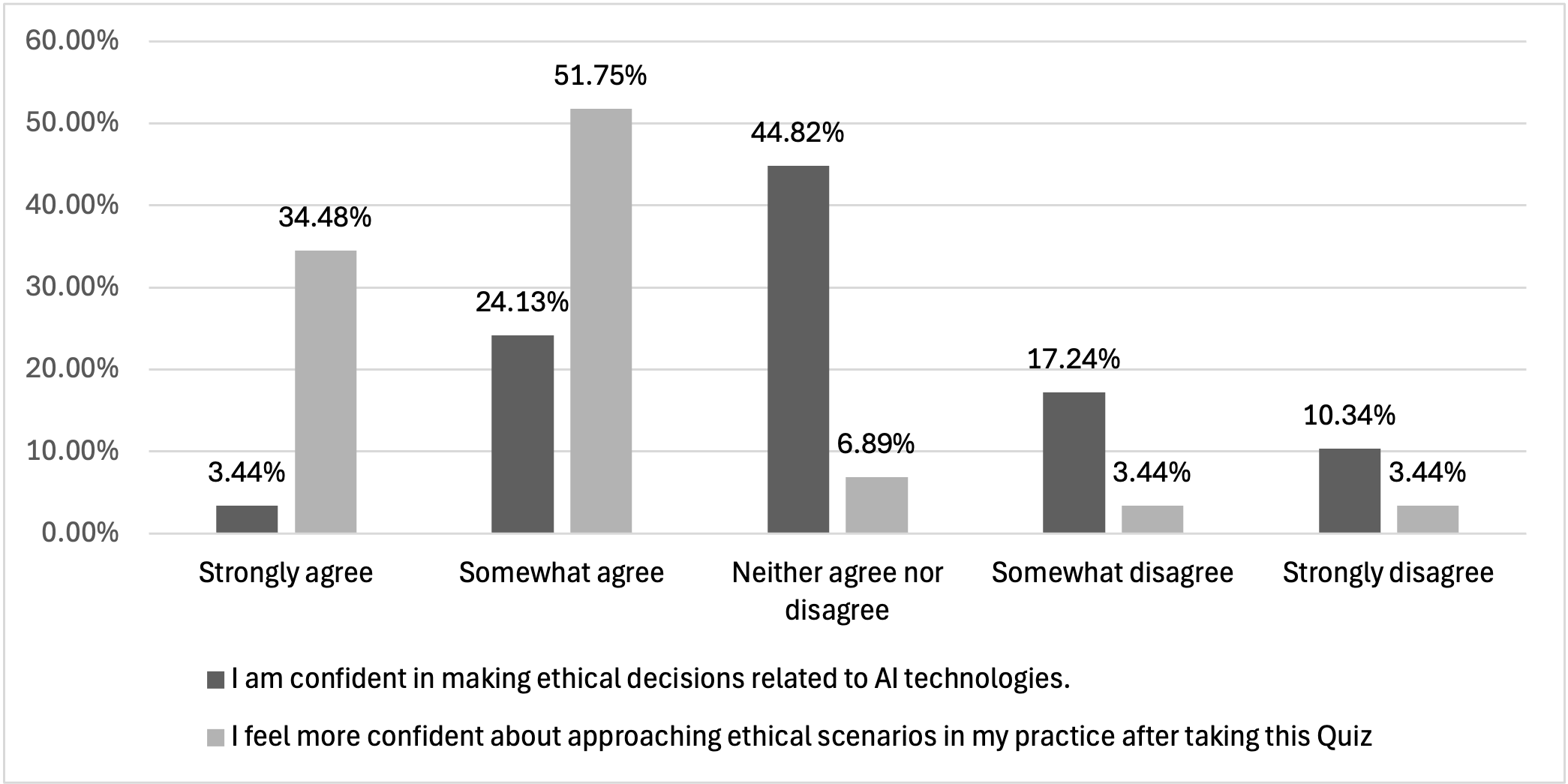}
    \caption{Practitioner responses about the efficacy of the quiz in improving confidence in approaching AI ethical scenarios (pre vs post-quiz)}
    \label{fig:Confidence}
\end{figure}

\par A majority (27.57\%) of practitioners agreed or strongly agreed that they were confident in making ethical decisions related to AI before taking the \emph{AI Ethics Quiz}, as shown in Figure \ref{fig:Confidence}. Meanwhile, 44.82\% neither agreed nor disagreed with this statement, and 27.58\% disagreed or strongly disagreed. After taking the quiz, 86.23\% of practitioners agreed or strongly agreed that it made them more confident in approaching ethical scenarios in their practice, with only 6.88\% disagreeing or strongly disagreeing. Only 6.89\% of the practitioners neither agreed nor disagreed with the statement. This represents a significant increase of 58.66 percentage points in practitioners' confidence in approaching AI ethical scenarios after taking the quiz.

\subsubsection{Engagement and usability of the quiz}
\textbf{RQ5:} How engaging is the \emph{AI Ethics Quiz} to learn about AI ethics?

\begin{figure}[htpb]
    \centering
    \includegraphics[scale=0.45]{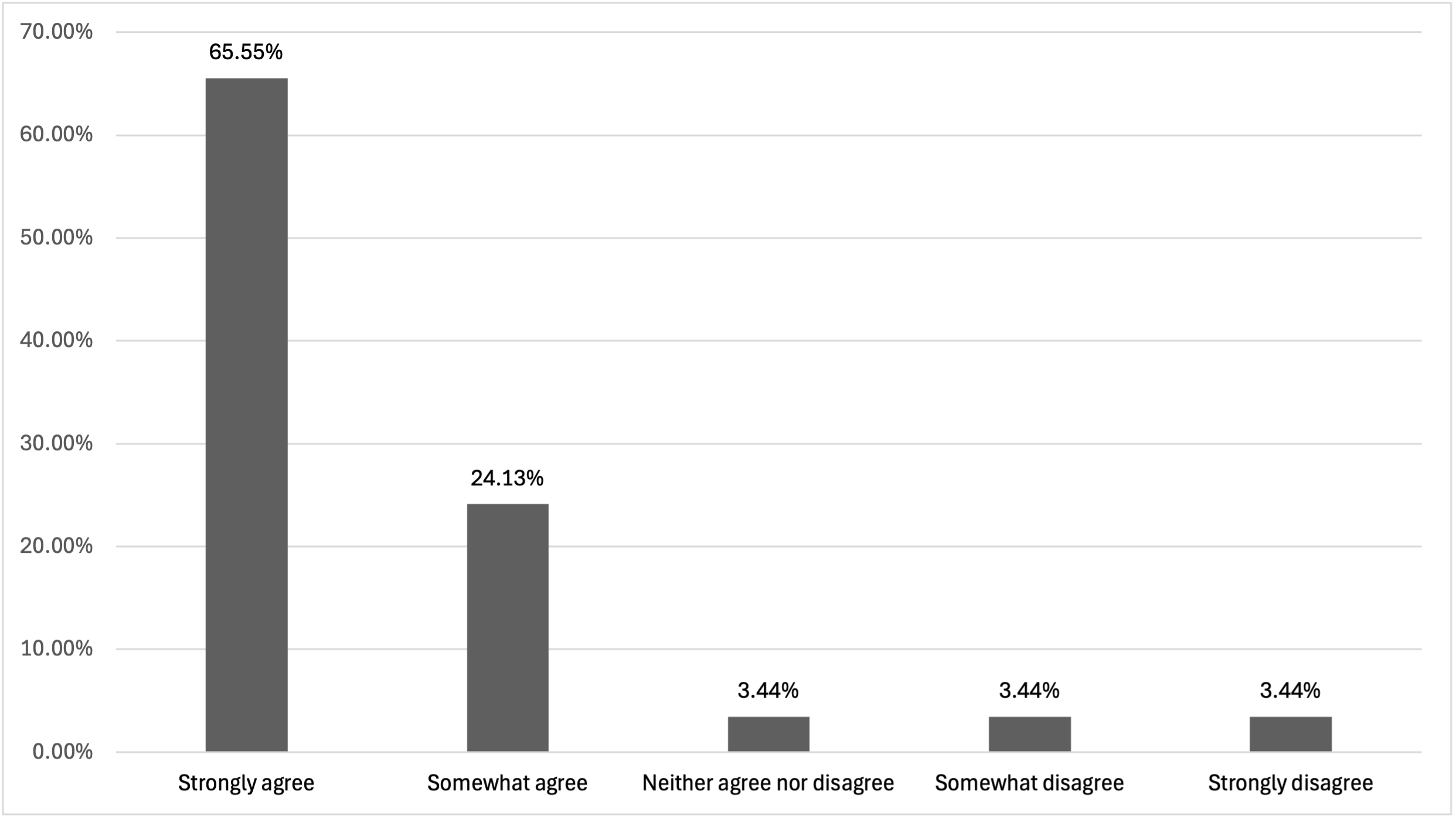}
    \caption{Practitioner responses about the engagement of the quiz (post-quiz)}
    \label{fig:Engagement}
\end{figure}

\par A majority (89.68\%) of practitioners agreed or strongly agreed that the \emph{AI Ethics Quiz} was an engaging way to learn about AI ethics, as shown in Figure \ref{fig:Engagement}. While 3.44\% of practitioners neither agreed nor disagreed with that, 6.88\% of practitioners disagreed or strongly disagreed that the quiz was engaging in learning about AI ethics. Overall, the data suggests that the quiz was well-received by most practitioners, though there was a small group that did not find it engaging.

\textbf{RQ6:} How difficult is it to find the ideal answers to the questions in the \emph{AI Ethics Quiz}?

\begin{figure}[htpb]
    \centering
    \includegraphics[scale=0.45]{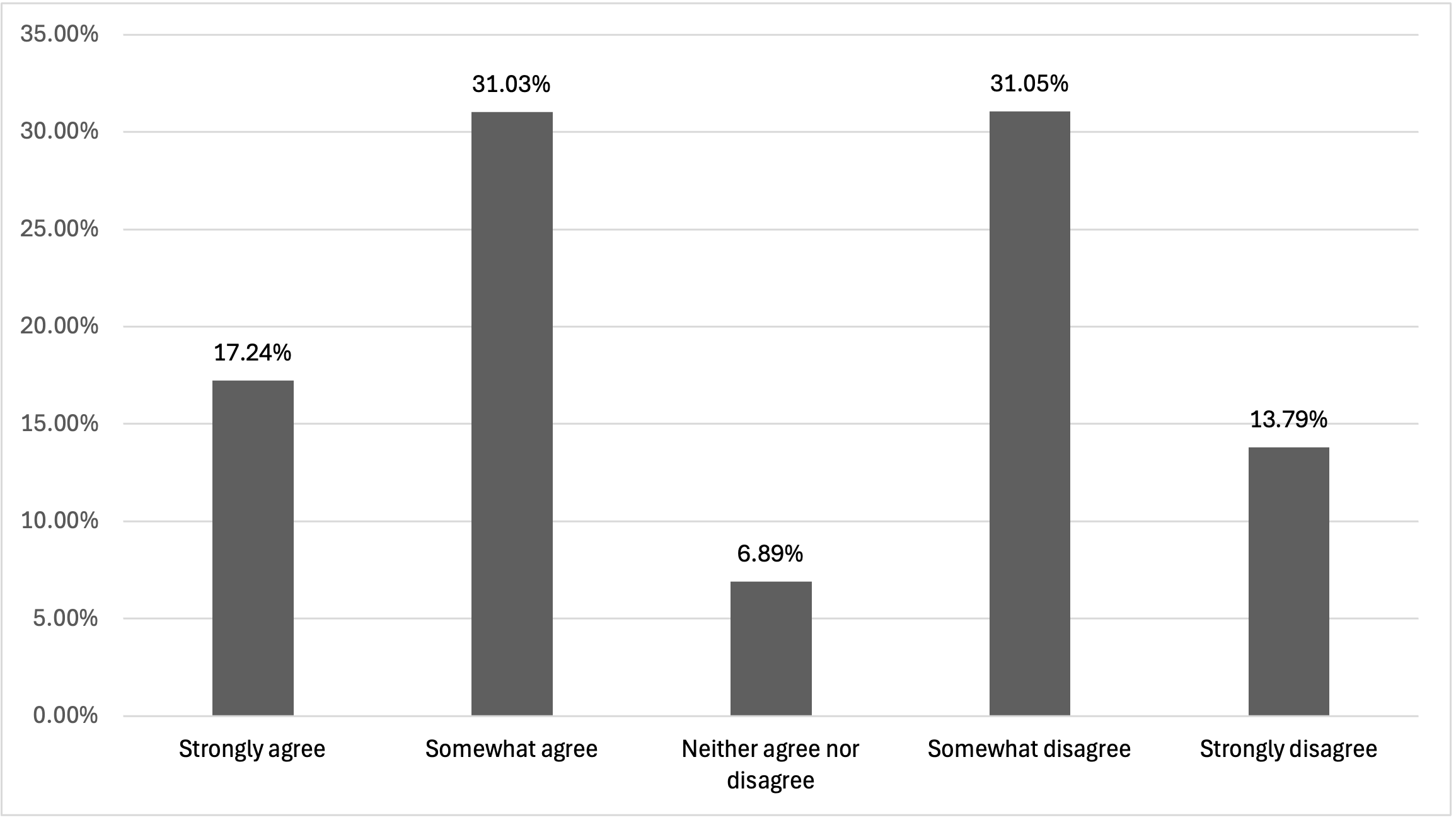}
    \caption{Practitioner responses about finding it hard to identify ideal answers (post-quiz)}
    \label{fig:Ideal_answer}
\end{figure}

\par The practitioners' perceptions of the difficulty in finding the ideal answers to the \emph{AI Ethics Quiz} were varied. Of the participants, 48.27\% agreed or strongly agreed that it was challenging to identify the ideal answers, while 44.84\% disagreed or strongly disagreed. Additionally, 6.89\% neither agreed nor disagreed with the statement, as shown in Figure \ref{fig:Ideal_answer}. The proportion of participants who found it difficult was slightly higher by 3.43 percentage points compared to those who did not. This finding addresses our RQ6 and demonstrates that individuals' perceptions of the difficulty in finding the ideal answers to the quiz questions differ.

\textbf{RQ7:} How well-integrated are the \emph{AI Ethics Quiz} scenarios, questions, and answer options to create a meaningful learning experience about AI ethics?

\begin{figure}[htpb]
    \centering
    \includegraphics[scale=0.45]{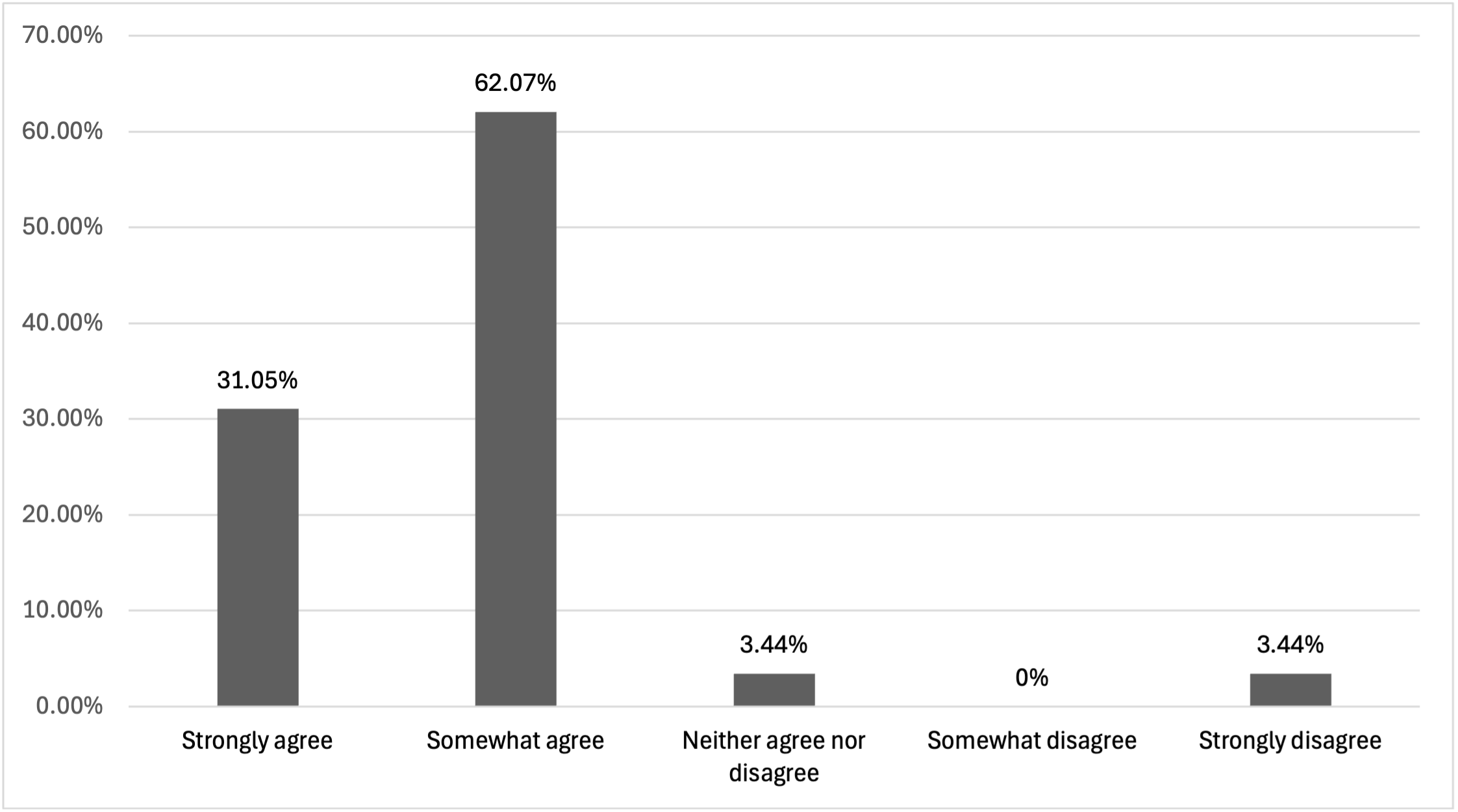}
    \caption{Practitioner responses on the integration of scenarios, questions, and answer options in the quiz (post-quiz)}
    \label{fig:Integration}
\end{figure}

When asked about the integration of the scenarios, questions, and answer options in the \emph{AI Ethics Quiz}, the majority of practitioners (93.12\%) perceived them to be well-integrated, as shown in Figure \ref{fig:Integration}. A smaller group, 3.44\%, neither agreed nor disagreed, while another 3.44\% strongly disagreed. Based on the overall data, we were able to answer our RQ7 as the majority of practitioners felt that the quiz's scenarios, questions, and answers were well-integrated and it created a meaningful learning experience about AI ethics. 

\section{Discussion} \label{sec:Discussion}
Our study aimed to introduce the \emph{AI Ethics Quiz} to software practitioners, raising their awareness of AI ethics and ethical principles while enhancing their understanding. We conducted two one-hour workshops—one in-person and one online—to achieve this. This section provides a summary of the quantitative findings presented earlier, a summary of the qualitative findings obtained through an open-text question, an analysis of the limitations and threats to the validity of our study, and lessons learned, along with recommendations and directions for future work.  

\subsection{Summary of Quantitative Findings}
\par Our results indicate that the software practitioners who participated in our study had a strong grasp of AI ethics and ethical principles, achieving an average score of 46.17 out of 52 (88.78\%). Notably, one practitioner achieved a perfect score of 52 out of 52 (100\%), while another scored the lowest at 30 out of 52 (58\%). With the exception of this one practitioner, all other participants scored above 75\%. Based on descriptive statistics, we found that overall, the practitioners had a good awareness of the concept of AI ethics before they took the \emph{AI Ethics Quiz}, as 72.41\% reported that they were aware of it. However, after taking the quiz, a total of 93.13\% reported that they are aware of the concept of AI ethics, which means that the \textbf{awareness} of AI ethics among practitioners increased by 20.72\% points after taking the quiz. It indicates that \emph{AI Ethics Quiz} is useful in raising awareness of AI ethics among software practitioners. The \emph{AI Ethics Quiz} has proven to be effective in raising awareness of AI ethics among software practitioners. However, the introductory talk on AI ethics and ethical principles that preceded the quiz during the workshop might have also contributed significantly to enhancing their awareness. Therefore, we recommend combining a talk on AI ethics with the \emph{AI Ethics Quiz} as a more effective approach to raising awareness of AI ethics among software practitioners. Similarly, our findings indicate a significant improvement in software practitioners' \textbf{knowledge and understanding} of AI ethics, with an increase of 41.39\% after taking the \emph{AI Ethics Quiz}. This substantial growth suggests that the quiz was highly effective in enhancing their understanding of AI ethics. While the accompanying talk on `AI ethics' may have contributed to this improvement, the notable 41.39 percentage point increase underscores the great impact of the \emph{AI Ethics Quiz} on their understanding and knowledge of AI ethics. 

\par Our results indicate that before taking the \emph{AI Ethics Quiz}, only 27.57\% of the practitioners in our study reported \textbf{confidence} in making ethical decisions. However, after completing the quiz, 86.23\% of the practitioners agreed that it boosted their confidence in addressing AI ethical scenarios in practice. This significant increase of 58.66 percentage points demonstrates that the quiz is highly effective in enhancing software practitioners' confidence in navigating ethical scenarios in AI. The scenario-based nature of the quiz likely contributes to this improvement, as it provides a practical exercise that encourages practitioners to think critically and ethically about each question.

\par In terms of the engagement and usability of the \emph{AI Ethics Quiz}, the majority of the practitioners, 89.68\%, reported that the \emph{AI Ethics Quiz} was \textbf{engaging} and effective for learning about AI ethics, while only 6.88\% disagreed. This finding suggests that our quiz successfully captured the interest of the participants and provided an engaging learning experience on AI ethics. Likewise, the findings reveal a slight difference of 3.43 percentage points in the practitioners' perception of the \textbf{difficulty in finding ideal answers} to the quiz questions. Specifically, 48.27\% reported that it was difficult to find ideal answers, while 44.84\% did not. This suggests that the difficulty level of the quiz varies depending on the individual. It can be concluded that the team’s effort in designing the quiz to be sufficiently challenging, but not overwhelming, was successful to an extent. In a similar vein, our findings show that the majority of the practitioners i.e. 93.12\% found the scenarios of the quiz, questions, and answer options very \textbf{well-integrated}. This implies that our quiz was successful in creating a meaningful learning experience for software practitioners. 

\par Overall, the findings indicate that our \emph{AI Ethics Quiz} effectively raises awareness of AI ethics among practitioners. It enhances their knowledge and understanding of AI ethical principles and boosts their confidence in addressing AI ethical scenarios in practice. Regarding engagement, most practitioners found the quiz to be highly engaging. The quiz is well-constructed, with scenarios, questions, and answer options that are thoughtfully integrated. It maintains a balanced level of difficulty, ensuring that the questions are neither too easy nor too difficult, facilitating a meaningful learning experience.

\subsection{Summary of Qualitative Findings}
In addition to the Likert-Scale questions, the post-quiz questionnaire form included one open-ended question which was:
\begin{itemize}
    \item Please share any other feedback or suggestions you have about the \emph{AI Ethics Quiz}.
\end{itemize}

\subsubsection{Feedback on AI Ethics Quiz}
To gather feedback and suggestions for future improvements, we asked software practitioners about their experience with the quiz. Out of 29 participants, 19 provided valuable input. We obtained \textbf{general} comments and feedback on the quiz. For instance, P1 said, \faCommenting \hspace{0.05cm}\emph{``I liked it a lot."}, P4 said, \faCommenting \hspace{0.05cm}\emph{``Very useful."}, and P16 described it as \faCommenting \hspace{0.05cm}\emph{``Very good."} However, these comments did not provide specific details about the aspects of the \emph{AI Ethics Quiz} that were appreciated.

\par Most practitioners commented on the \textbf{interactive nature} of the quiz and its \textbf{engagement}. For example, P7 said, \faCommenting \hspace{0.05cm}\emph{``It was an engaging way to trigger the thought process"}, whereas, P10 said, \faCommenting \hspace{0.05cm}\emph{``The quiz was really nice and engaging. Thank you for this session"}. Similarly, practitioners appreciated that the quiz encouraged reflection and \textbf{stimulated their thoughts} on AI ethics. For instance, P6 noted, \faCommenting \hspace{0.05cm}\emph{``Was a great interactive quiz which was thought-provoking and used close to real-life examples."},  whereas, P18 said, \faCommenting \hspace{0.05cm}\emph{``Was a good experience, made me think"} and P24 said, \faCommenting \hspace{0.05cm}\emph{``I thought it was a good way to get us thinking about the different scenarios around AI ethics."} Additionally, practitioners also provided feedback on the \textbf{ease} of taking the \emph{AI Ethics Quiz}. For instance, P9 said, \faCommenting \hspace{0.05cm}\emph{``The quiz was good and engaging, also quite easy to go through."} and P13 said, \faCommenting \hspace{0.05cm}\emph{``It was easy to follow."} A practitioner, P26 also explicitly mentioned that taking the quiz enhanced their \textbf{awareness} of AI ethics, stating, \faCommenting \hspace{0.05cm}\emph{``Thanks for opening my eyes and making me more aware of AI ethics."} These positive comments and feedback from software practitioners regarding their experience with the \emph{AI Ethics Quiz} indicate that the quiz is interactive, engaging, easy to follow, and helps in raising awareness of AI ethics which contributes to a more meaningful and effective learning experience in AI ethics for practitioners. 

\par Some practitioners also commented on the nature of the \textbf{quiz scenarios and questions}. For instance, P15 noted the balance between answer options that reflect purely ethical considerations and those influenced by commercial factors, stating, \faCommenting \hspace{0.05cm}\emph{``It interestingly highlights the difficulty that occurs between the ``pure" ethical response and an ethical response that may be tainted by commercial considerations."} However, one participant, P17, suggested that the quiz could benefit from more challenging questions, commenting, \faCommenting \hspace{0.05cm}\emph{``The quiz was really good but I feel like it should have some harder questions in there that test peoples' priorities."}

\par While most practitioners praised the quiz for its interactive feature, engagement, and ease of use, one participant found the feedback provided based on scoring to be unhelpful. Participant P22 commented: \faCommenting \hspace{0.05cm}\emph{``Ethics is inherently subjective, measuring the results of a subjective subject with an objective rubric was off-putting, even with the disclaimers of `there are grey areas / no right answer' especially since it concluded with being a feedback message depending on your score."} We introduced our \emph{AI Ethics Quiz} \cite{teo2023would} to software practitioners, which included hypothetical scenarios, questions based on those scenarios, and four answer options for each question. These options were labeled as `least desirable’, `bearable’, `less than ideal’, and `ideal’, corresponding to scores of 1, 2, 3, and 4, respectively. Although the final score was not directly visible to users, the quiz concluded with a \emph{Result Summary} page that provided feedback on their overall performance. Participant P22 expressed that ethics is inherently subjective, based on personal beliefs and values, and therefore, evaluating it using an objective rubric (like the \emph{AI Ethics Quiz}) can feel inappropriate. We somewhat agree with this statement. \emph{Ethics} is indeed a grey area; what is considered ethical by one person may not be viewed the same way by another. However, during the design of the quiz, our team of ten people engaged in multiple meetings and discussions to label the answer options as `least desirable’, `bearable’, `less than ideal’, and `ideal’. These labels were not based on one or two individuals' opinions but were agreed upon by all ten team members. This collaborative approach ensured that the answer options reflected a broader consensus rather than individual viewpoints. Detailed information about the design and development of the \emph{AI Ethics Quiz} can be found in our other paper by \citet{teo2023would}.  
Similarly, the primary goal of this quiz was to raise awareness of AI ethics among practitioners and help them self-assess their knowledge and understanding. Although we included feedback at the end of the quiz to give it a quiz-like vibe, the main purpose was to create a fun, interactive, and engaging exercise to enhance users' awareness and understanding of AI ethics.

\subsubsection{Feedback on the workshop session}
\par In addition to the \emph{AI Ethics Quiz}, some practitioners also shared their feedback on the workshop session. All practitioners felt that the workshop was beneficial and provided positive feedback. For instance, P18 remarked, \faCommenting \hspace{0.05cm} \emph{``The session was good."} Meanwhile, P29 suggested increasing the visibility and publicity of the workshop to attract a larger audience, stating, \faCommenting \hspace{0.05cm} \emph{``This session could be a lot more publicised. Not a lot [of] visibility."} Participant P20 expressed an interest in learning more about regulatory compliance and laws in Australia related to AI, commenting, \faCommenting \hspace{0.05cm} \emph{``Well construed presentation. Would have loved to know more about regulatory compliance and laws in Australia."} Similarly, participant P3 suggested providing follow-up materials after the workshop for further research on the discussed topics, stating, \faCommenting \hspace{0.05cm} \emph{``Provide details of follow-up material to do further research in the topic."} In response, we provided the presentation slides and added all relevant links to the slides after the workshop, allowing practitioners to explore the topics in greater depth if they were interested. Overall, the feedback on the workshop sessions was overwhelmingly positive, indicating that organising such workshops could be an effective way to raise awareness of AI ethics among software practitioners.

\subsection{Limitations and Threats to Validity}
We acknowledge several limitations and potential threats to the validity of our study, which are discussed here. The first limitation of our study is that it is based on the responses from only 29 participants and a single IT company based in Australia. This limits the generalisability of our findings, as they are specific to the practitioners of this particular company and may not be representative of a larger population of software practitioners. Additionally, while our quiz is based on Australia's AI Ethics Principles \cite{Australisethics}, it does not encompass all the principles. Instead, it focuses on four key ethical principles of AI: \emph{accountability}, \emph{fairness}, \emph{privacy}, and \emph{human, societal \& environmental well-being}. This was intentional to avoid overloading the quiz with too many scenarios and questions. Moreover, designing scenarios and questions based on the other principles posed significant challenges, which we plan to address in future studies. Similarly, the findings of our study are derived from responses by software practitioners, many of whom had an interest in AI but limited experience in AI development. Thus, testing with experienced AI practitioners may show different results. However, since software practitioners are increasingly becoming exposed to AI applications, tools, and system development, such quizzes can be an effective and engaging way to initiate discussions on \emph{AI ethics}. Additionally, this lack of experience may have influenced certain findings, such as their level of awareness and knowledge/ understanding of AI ethics, which is a limitation of our study.  

\par Given that we conducted two workshops—one in-person and one online, each lasting one hour—there is a possibility of discrepancies in our study's findings. For example, participants in the in-person workshop might have gained a better understanding of AI ethics concepts compared to those who joined online as in-person workshops tend to be more interactive, facilitating better engagement and comprehension than online workshops \cite{kim2022comparing}. This may have influenced their rating of the question regarding the effectiveness of the \emph{AI Ethics Quiz} in enhancing their knowledge and understanding of AI ethics, as discussed in Section \ref{sec:RQ3}. 

\par To protect participants' privacy and avoid conflicts of interest—since one of the authors of this paper was both the workshop organiser and one of the leaders of the company's IT team —the data collection process through the pre/post quiz questionnaire and the \emph{AI Ethics Quiz} was made completely anonymous. Participants in the in-person workshop were assigned a unique ID by asking them to pick a random number from a bowl, while those in the online workshop selected a number between 300 and 700 to serve as their ID. To prevent duplicates in the online workshop, we ensured that no two participants chose the same number, and no two participants did that. This approach helped mitigate potential threats to the validity of our study.

\subsection{Lessons and Recommendations}
Based on our experience conducting AI ethics workshops and collecting the data using our \emph{AI Ethics Quiz}, we offer key lessons learned and recommendations for software companies and practitioners.\\

\noindent \faThumbsUp~ \textbf{Combining Talks and Quizzes for Impact:} We conducted a one-hour workshop twice with software practitioners: one session in-person and another online. During these workshops, we began by delivering a talk on \emph{AI ethics}, emphasising its importance in today's context, introducing Australia's AI Ethics Principles \cite{Australisethics}, and exploring various approaches to AI ethics. The feedback we received highlighted that the talk on AI ethics was particularly beneficial for practitioners. Even though we asked the practitioners for their feedback on improving the quiz, we obtained the feedback primarily about the talk. To enhance the learning experience for practitioners, we recommend incorporating such talks alongside the \emph{AI Ethics Quiz} by software companies. These talks provide essential context, reducing the risk of practitioners feeling overwhelmed by the quiz alone, as relying solely on talks may be ineffective \cite{bairaktarova2015engineering}. Additionally, inviting academics, professionals, and policymakers to give talks on AI ethics can further enrich the learning experience and deepen practitioners' understanding of this crucial topic as similar approaches have been used in the workshops on AI ethics \cite{bendechache2021ai}.

\noindent \faThumbsUp~ \textbf{Broadening AI Ethics Talks:} During our talk on \emph{AI Ethics}, we provided an overview of foundational aspects, including an introduction to AI ethics and Australia's AI Ethics Principles \cite{Australisethics}. However, feedback indicated a need for more detailed coverage of compliance and legal frameworks specific to AI technologies. To address this, we will consider expanding the talk to include a more comprehensive exploration of AI ethics. This could also involve discussing global ethical principles, such as those established by the European Commission \cite{EUethics}, and those introduced by major tech companies like Microsoft \cite{Microsoft}, Google \cite{Google}, and IBM \cite{IBM}. By incorporating these elements, the talk would offer a broader perspective and deeper understanding of AI ethics, helping practitioners better grasp the various principles and frameworks that shape ethical AI practices worldwide.

\noindent \faThumbsUp~ \textbf{Allocating additional time for workshop:} We allocated one hour for the workshop, but the schedule proved tight. Of the hour, 20 minutes were dedicated to the talk, 15 minutes to the \emph{AI Ethics Quiz}, 10 minutes each for pre- and post-quiz questionnaires, and the remaining time for a collaborative debrief session to discuss quiz answers. Due to time constraints, we slightly exceeded the allotted hour, causing some participants to leave before completing the post-quiz questionnaire. As a result, we had to exclude responses from those who completed the pre-quiz and \emph{AI Ethics Quiz} but not the post-quiz questionnaire. This experience highlighted the need for additional time in such workshops, as discussions and varying answers during the debrief session can extend beyond expectations. We recommend allocating extra time to accommodate these factors and ensure all participants can complete the workshop components.

\noindent \faThumbsUp~ \textbf{Respecting other's viewpoints:} \emph{Ethics} is inherently complex and often subjective. Our experiences have shown that navigating ethical scenarios requires a balanced approach, and it can be quite challenging. For example, during our collaborative debriefing session, we discussed the ideal responses to various quiz questions. We then asked participants to justify their choices if they had selected answers other than the ideal ones. This revealed a range of viewpoints among participants, with not everyone agreeing on the ideal response. Based on these experiences, we recommend reminding participants that ethical judgments can be subjective and that respecting differing perceptions is crucial. Additionally, we suggest designing quiz questions with a mix of simpler and more complex ethical dilemmas. Research indicates that a lack of contextual variety in questions can make responses less reliable as predictors of behavior, potentially undermining the effectiveness of ethics training \cite{blatner2009role}.

\subsection{Future Work}
The feedback obtained from the post-quiz questionnaire provided valuable insights for future work. In this study, we conducted a one-hour workshop twice with software practitioners, focusing on AI ethics. During the workshop, we introduced our \emph{AI Ethics Quiz} to gather feedback on its effectiveness in enhancing participants' awareness and understanding of AI ethics. Participants included both those with and without experience in developing AI-based systems. For future studies, we propose conducting a proper intervention study with AI practitioners who have experience in AI development activities to evaluate the effectiveness of the \emph{AI Ethics Quiz} in raising their awareness and understanding of AI ethics. Additionally, one of the feedback points from the post-quiz questionnaire suggested increasing the publicity of the workshop sessions. Previously, we advertised the workshop only within one company through an internal system. In the future, we plan to use social media platforms such as LinkedIn and Twitter to attract more participants as these workshops proved to be beneficial for software practitioners in learning about AI ethics. 

\par Moreover, our \emph{AI Ethics Quiz} is a web application that is not mobile-friendly and can only be used on a laptop. Participants in the in-person workshop had to bring their laptops to take the quiz, although they could use their mobile phones for other surveys like the pre-quiz and post-quiz questionnaires. We received feedback from the in-person workshop that participants would prefer to take the quiz using their mobile phones. Therefore, we plan to make our quiz mobile-friendly in future iterations.

\section{Conclusion} \label{sec:Conclusion}

The recent ethical challenges associated with AI-based systems highlight the critical need for raising awareness and understanding of AI ethics among software practitioners. The purpose of this paper is to present our experience with introducing the \emph{AI Ethics Quiz} to software practitioners to enhance awareness and understanding of AI ethics. To achieve this, we conducted a one-hour workshop twice—once in person and once online with software practitioners of an Australian-based IT company. The workshop began with a presentation on `AI ethics', including a discussion on Australia's AI Ethics Principles \cite{Australisethics}, and the significance of AI ethical considerations in today's time. We collected data in three stages: (i) a pre-quiz questionnaire, (ii) the \emph{AI Ethics Quiz}, and (iii) a post-quiz questionnaire. Our goal was to gauge the effectiveness of the quiz in raising awareness and improving knowledge and understanding of AI ethics among software practitioners. A total of 29 participants completed all three stages of data collection and their participation was voluntary and anonymous.

\par We found that participants in our study had a good understanding of AI ethics as most of them scored over 75\% in the quiz. Taking the quiz increased the confidence of software practitioners in approaching ethical scenarios of AI greatly, and they found the quiz highly engaging. However, there was mixed feedback regarding the difficulty of finding ideal answers to some questions, though the findings indicate that the participants found the scenarios, quiz questions, and answer options well-integrated. Overall, our quiz successfully contributed to increasing awareness and understanding of AI ethics among software practitioners. We also share insights and recommendations based on our experience conducting the workshop, aiming to encourage software companies and leaders to adopt similar initiatives. We hope our lessons and recommendations will inspire further efforts to promote AI ethics awareness within the software industry.\\

\textbf{CRediT authorship contribution statement:}
\textbf{Aastha Pant:} Writing– original draft, Visualisation, Project administration, Methodology, Formal analysis, Data curation, Conceptualisation. \textbf{Rashina Hoda:} Writing- review \& editing, Validation, Supervision, Methodology. \textbf{Paul McIntosh:} Writing- review \& editing, Validation, Supervision, Methodology.

\textbf{Declaration of competing interest:}
The authors declare the following financial interests/personal relationships which may be considered as potential competing interests: Aastha Pant reports financial support was provided by the Faculty of Information Technology Ph.D. scholarship from Monash University. 

\textbf{Acknowledgement:}
Aastha Pant is supported by the Faculty of IT Ph.D. scholarship from Monash University.

\textbf{Data availability}
The data are protected and are not available due to data privacy laws.

\appendix
\section{Demographic Questions} \label{appendix1}
\scriptsize
 \begin{enumerate}[1]
     \item How old are you?
     \begin{itemize}
         \item Below 20
\item 20-25
\item  26-30
\item 31-35
\item 36-40
\item 41-45
\item 46-50
\item Above 50
     \end{itemize}

     \item How would you describe your gender?
     \begin{itemize}
         \item Man
\item Woman
\item Prefer to self-describe as:
\item  Prefer not to answer
     \end{itemize}

     \item What is the highest degree or level of education you have completed?
     \begin{itemize}
         \item High School
\item Bachelor’s degree
\item Master’s degree
\item  Ph.D. or Higher
\item  Prefer not to answer
\item  Others:
     \end{itemize}

     \item What is your current job title?
     \begin{itemize}
         \item AI Engineer
\item AI/ Data Scientist
\item  AI/ML Specialist
\item AI Expert
\item  AI/ML Practitioner
\item  AI Developer
\item Prefer not to answer
\item Others: 
     \end{itemize}

     \item How many years of experience do you have in AI-based software development?
     \begin{itemize}
         \item No experience
\item Less than 1 year
\item Between 1 to 2 years
\item Between 3 to 5 years
\item Between 6 to 10 years
\item Between 11-15 years
\item Between 16-20 years
\item Over 20 years
     \end{itemize}
 \end{enumerate}

\normalsize




\bibliographystyle{plainnat} 
\bibliography{reference}



\end{document}